\documentclass[twocolumn,aps,prl,preprintnumbers]{revtex4-2}
\usepackage{bbm}
\usepackage[latin9]{inputenc}
\usepackage{amsmath}
\usepackage{amssymb}
\usepackage{graphicx}
\usepackage{mathrsfs}
\usepackage{amsfonts}
\usepackage{bm}
\usepackage{amsthm}
\usepackage{color}
\usepackage{txfonts}
\usepackage[colorlinks=true,citecolor=blue,linkcolor=blue,urlcolor=blue,anchorcolor=blue]{hyperref}%
\usepackage{pifont}
\hypersetup{colorlinks=true,citecolor=blue,linkcolor=blue,urlcolor=blue}
\providecommand{\U}[1]{\protect\rule{.1in}{.1in}}
\makeatletter
\@ifundefined{textcolor}{}
{
\definecolor{BLACK}{gray}{0}
\definecolor{WHITE}{gray}{1}
\definecolor{RED}{rgb}{1,0,0}
\definecolor{GREEN}{rgb}{0,1,0}
\definecolor{BLUE}{rgb}{0,0,1}
\definecolor{CYAN}{cmyk}{1,0,0,0}
\definecolor{MAGENTA}{cmyk}{0,1,0,0}
\definecolor{YELLOW}{cmyk}{0,0,1,0}
}

\makeatother
\begin{document}
\title{Interaction-Driven Altermagnetic Magnon Chiral Splitting}
\author{Zhejunyu Jin$^1$}
\author{Zhaozhuo Zeng$^1$}
\author{Jie Liu$^1$}
\author{Tianci Gong$^1$}
\author{Ying Su$^1$}
\author{Kai Chang$^2$}
\email[Contact author: ]{kchang@zju.edu.cn}
\author{Peng Yan$^{1,3}$}
\email[Contact author: ]{yan@uestc.edu.cn}
\affiliation{$^1$School of Physics and State Key Laboratory of Electronic Thin Films and Integrated Devices, University of Electronic Science and Technology of China, Chengdu 610054, China\\
$^2$Center for Quantum Matter, School of Physics, Zhejiang University, Hangzhou 310058, China\\
$^3$Institute of Fundamental and Frontier Sciences, Key Laboratory of Quantum Physics and Photonic Quantum Information of the Ministry of Education, University of Electronic Science and Technology of China, Chengdu 611731, China}

\begin{abstract}
Nonrelativistic magnon chiral splitting in altermagnets has garnered significant recent attention. In this work, we demonstrate that nonlinear three-wave mixing---where magnons split or coalesce---extends this phenomenon into unprecedented relativistic regimes. Employing a bilayer antiferromagnet with Dzyaloshinskii-Moriya interactions, we identify three distinct classes of chiral splitting, each dictated by specific symmetries, such as $C_4T$, $\sigma_v T$, or their combination. This reveals a novel bosonic mechanism for symmetry-protected chiral splitting, capitalizing on the unique ability of magnons to violate particle-number conservation, a feature absent in low-energy fermionic systems. Our findings pave the way for  engineering altermagnetic splitting, with potential applications in advanced magnonic devices and deeper insights into magnon dynamics in complex magnetic systems.
\end{abstract}

\maketitle

\textit{Introduction---}Altermagnets (ATMs) represent an emerging class of magnetic materials that redefine and expand our understanding of antiferromagnetism. Unlike conventional antiferromagnets, ATMs exhibit compensated magnetic moments that yield zero net magnetization, yet they violate the combined parity-time ($PT$) symmetry while preserving specific rotational symmetries in their spin and crystal structures \cite{Hayami2019,Hayami2020,Smejkal1,Smejkal3,McClarty2024,Gomonay2024,Jin2024,Duan2025}. 
This unique symmetry profile induces a remarkable, momentum-dependent spin-splitting effect in their electronic band structures \cite{Duan2025,Osumi2024,Jiang2025,Reimers2024,Lee2024,Ding2024,Song2025}. 
Initial studies of ATMs centered on simplified non-relativistic models without spin-orbit coupling (SOC), but recent advancements have explored more complex regimes, integrating relativistic effects and electron-electron interactions \cite{Sato2024}. For instance, in a modified Kane-Mele model with broken inversion symmetry, the spin splitting can be stabilized by electron interactions \cite{Sato2024}. These developments highlight the need to move beyond single-particle descriptions toward the more comprehensive framework of many-body physics to fully capture the intricacies of altermagnets.
\begin{figure}[t]
  \centering
  \includegraphics[width=0.48\textwidth]{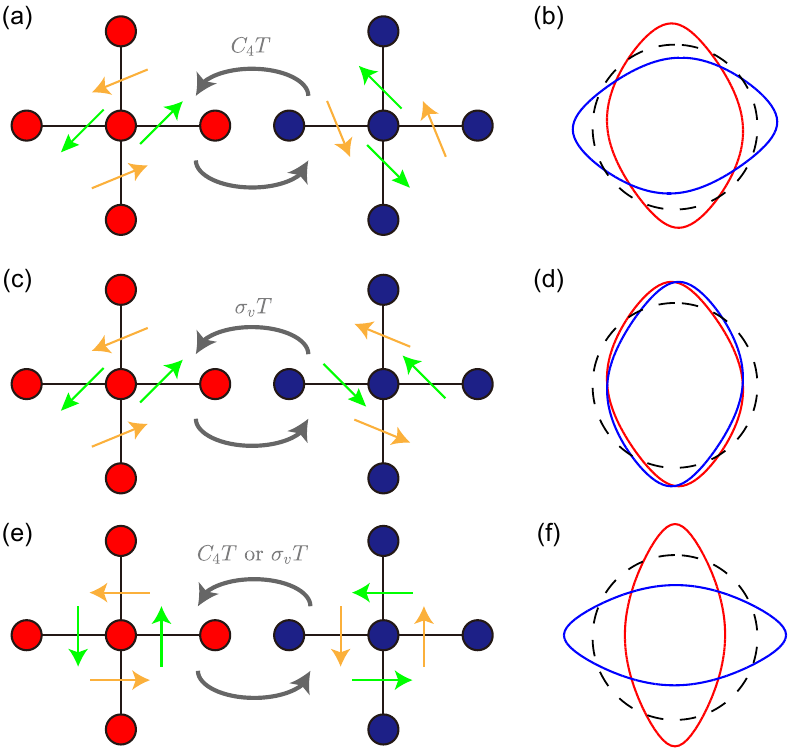}\\
  \caption{Schematic illustrations of DMI-induced altermagnetic chiral splitting in relativistic antiferromagnets under different symmetries. (a, c, e) DMI configurations for Hamiltonians $H^{\rm C}$, $H^{\rm M}$, and $H^{\rm CM}$, preserving $C_4 T$, $\sigma_v T$, and both symmetries, respectively. Red and blue circles represent spins with opposite orientations, while green and orange arrows denote DM vectors of differing magnitudes. (b, d, f) Corresponding magnon spectra for opposite chiralities, linked by $C_4$, $\sigma_v$, and $C_4$ or $\sigma_v$ operations, respectively. Red and blue curves show isoenergy contours for right- and left-handed magnons as derived from anharmonic theory, with dashed black lines indicating degenerate contours from LSWT.}\label{fig1}
\end{figure}

Magnons, the quanta of spin waves, offer tremendous potential for advancing magnonics and spintronics. Their capacity to transmit information over long distances with low energy dissipation positions them as essential components in ultra-efficient data processing technologies \cite{Kruglyak2010,Chumak2015,Pirro2021,Yuan2022}. In ATMs, magnons exhibit a distinctive chiral splitting, analogous to the spin splitting in electronic band structures \cite{Smejkal2,Liu2024,Hoyer2025}. However, unlike electrons, magnons are bosons whose particle number is not conserved. This property facilitates nonlinear interactions, including three-wave mixing, in which magnons can split or coalesce \cite{Zhitomirsky2013,Wang2021}. Such nonlinearities can violate symmetries preserved in linear spin-wave theory (LSWT) \cite{McClarty2019,Chernyshev2009,Mook2021,Fujiwara2025,Gohlke2023}, paving the way for emergent physical phenomena \cite{Zheng2023}. A key question thus arises: Can the bosonic nature of magnons, via processes like three-wave mixing, induce altermagnetic chiral splitting in conventional antiferromagnets? Addressing this inquiry bridges the emerging field of altermagnetism with the complex dynamics of magnons, deepening our insights into the origins of chiral splitting.

In this Letter, we demonstrate that three-magnon interactions can induce altermagnetic chiral splitting in relativistic antiferromagnets. We examine a bilayer compensated magnet incorporating Dzyaloshinskii-Moriya interactions (DMIs) under various symmetries. Our analysis elucidates the critical role of DMIs in mediating nonlinear magnon-magnon interactions. Applying the second-order many-body perturbation theory, we derive the renormalized magnon spectrum shaped by these three-magnon processes. Strikingly, we reveal that DMIs generate an effective SOC, yielding symmetry-protected magnon chiral splitting governed by $C_4 T$, $\sigma_v T$, or their combination. This mechanism is inherent to the bosonic nature of magnons, where the particle-number is not conserved, in stark contrast to their fermionic counterparts. Our results advance the understanding of altermagnetism and pave the way for engineering tunable chiral splitting in magnon-based devices, with transformative potential for next-generation spintronics and quantum technologies.
 
\textit{Model---}We consider a bilayer compensated magnet with broken inversion symmetry modeled by the Hamiltonian
\begin{equation}\label{Eq1}
\begin{aligned}
{\mathcal H}^{\rm C/M/CM}&=-\sum_{{\bf r}, {\bm \delta}}\big[J({\bf S}_{A,{\bf r}} \cdot{\bf S}_{A,{\bf r}+{\bm \delta}}+{\bf S}_{B,{\bf r}} \cdot{\bf S}_{B,{\bf r}+{\bm \delta}})+J_0{\bf S}_{A,{\bf r}}\cdot{\bf S}_{B,{\bf r}}\\
&-{\bf D}_{A,\bm\delta}^{\rm C/M/CM}\cdot({\bf S}_{A,{\bf r}}\times{\bf S}_{A,{\bf r}+{\bm \delta}})-{\bf D}_{B,\bm\delta}^{\rm C/M/CM}\cdot({\bf S}_{B,{\bf r}}\times{\bf S}_{B,{\bf r}+{\bm \delta}})\\
&+K({\bf S}_{A,{\bf r}}\cdot {\bf z})^2+K({\bf S}_{B,{\bf r}}\cdot {\bf z})^2\big],
\end{aligned}
\end{equation}
where $J>0$ denotes the isotropic intralayer ferromagnetic exchange coupling, and $J_0<0$ represents the interlayer antiferromagnetic exchange constant. The in-plane DM vectors for sublattices $A$ and $B$ along the nearest-neighbor (NN) direction ${\bm \delta}$ are denoted as ${\bf D}_{A/B,\bm\delta}^{\rm C/M/CM}$. The NN vectors are ${\bm \delta}=a(0,1)$ and $a(1,0)$ with $a$ the lattice constant. The parameter $K$ represents the magnetic anisotropy, and ${\bf S}_{A,{\bf r}} $ and ${\bf S}_{B,{\bf r}}$ are spin operators on sublattices $A$ and $B$ at site ${\bf r}$ with spin length $S$, respectively. Unless otherwise specified, we set $J_0 = -J$, $S=1$, and $K=2J$ to prevent overlap between the harmonic single-particle energy and the two-magnon continuum \cite{Mook2021}.

To realize altermagnetic magnons, we examine three types of DMIs: ${\bf D}_{A/B,\bm\delta}^{\rm C}$, ${\bf D}_{A/B,\bm\delta}^{\rm M}$, and ${\bf D}_{A/B,\bm\delta}^{\rm CM}$, corresponding to four-fold crystallographic rotation symmetry ($C_4$), mirror reflection ($\sigma_v$), and their combination, respectively. These configurations are illustrated in Figs. \ref{fig1}(a), (c), and (e). Below, we analyze the symmetries to confirm that each DMI type respects its associated symmetry and yields a distinct $d$-wave altermagnetic chiral splitting, depicted in Figs. \ref{fig1}(b), (d), and (f).

Unlike non-relativistic antiferromagnets, which rely solely on spin space groups \cite{Chen2025}, those with SOC require coordinated transformations in both spin and crystallographic spaces \cite{Corticelli2022}. Under the $C_4$ operation, defined by $x\rightarrow-y$, $y\rightarrow x$, with spin transformations $S^x_{A(B)}\rightarrow-S^y_{A(B)}$, and $S^y_{A(B)}\rightarrow S^x_{A(B)}$, and time reversal $T$ (which maps ${\bf S}_A\rightarrow {\bf S}_B$), the Hamiltonians ${\mathcal H}^{\rm C}$ and ${\mathcal H}^{\rm CM}$ remain invariant. Consequently, these systems preserve $C_4 T$ symmetry, connecting magnon spectra of opposite chiralities via $C_4$, as shown in Figs. \ref{fig1}(b) and (f), and exhibiting $d$-wave altermagnetic chiral splitting. For mirror symmetry $\sigma_v$, we consider two representative planes. The $y-z$ plane acts as $x\rightarrow-x$, with $S^y_{A(B)}\rightarrow-S^y_{A(B)}$, and $S^z_{A(B)}\rightarrow -S^z_{A(B)}$. The $x-z$ plane maps $y\rightarrow-y$, $S^x_{A(B)}\rightarrow-S^x_{A(B)}$, and $S^z_{A(B)}\rightarrow -S^z_{A(B)}$. Combined with $T$, ${\mathcal H}^{\rm M}$ remains invariant, indicating protection by $\sigma_v T$ symmetry and yielding the chiral splitting pattern in Fig. \ref{fig1}(d). For ${\mathcal H}^{\rm CM}$, we further consider mirror planes at $45^\circ$ and $135^\circ$ relative to the $x-$axis, mapping $x\rightarrow \pm y$, $y\rightarrow \pm x$, with spins transforming as $S^x_{A(B)}\rightarrow\mp S^y_{A(B)}$, $S^y_{A(B)}\rightarrow\mp S^x_{A(B)}$, and $S^z_{A(B)}\rightarrow -S^z_{A(B)}$. Notably, ${\mathcal H}^{\rm CM}$ is invariant under either $C_4 T$ or $\sigma_v T$, ensuring that opposite-chirality magnon spectra are linked by either symmetry, as in Fig.\ref{fig1}(f). Further symmetry details are provided in Sec. I of Supplemental Material \cite{SM}. Table~\ref{table1} summarizes the symmetries that stabilize each DMI type's splitting. While this framework outlines the principles, the underlying physics proves more intricate, as explored next.
\begin{table}[t]
\centering
\caption{Symmetries stabilizing altermagnetic chiral splitting in the three models.}
\label{table1}
\begin{tabular}{c c c c}
\hline\hline
& ${\mathcal H}^{\rm C}$ [Fig. \ref{fig1}(a)] & ${\mathcal H}^{\rm M}$ [Fig. \ref{fig1}(c)]& ${\mathcal H}^{\rm CM}$ [Fig. \ref{fig1}(e)]\\
\hline
{\rm $C_4T$ symmetry} & \ding{52} & \ding{56} & \ding{52}\\
{\rm $\sigma_vT$ symmetry} & \ding{56} & \ding{52} & \ding{52}\\
\hline\hline
\end{tabular}
\end{table}

\textit{LSWT---}To study magnon excitations, we employ the Holstein-Primakoff transformation to express spin operators in terms of bosonic operators \cite{HP}. For sublattice A, we introduce creation and annihilation operators $a^{\dagger}$ and $a$, respectively; for sublattice B, we use $b^{\dagger}$ and $b$. Expanding the Hamiltonian to fourth order yields $\mathcal{H}=\mathcal{H}_2+\mathcal{H}_3+\mathcal{H}_4$, where $\mathcal{H}_2$, $\mathcal{H}_3$, and $\mathcal{H}_4$ denote quadratic (two-magnon), cubic (three-magnon), and quartic (four-magnon) terms, respectively. In LSWT, we retain only $\mathcal{H}_2$, describing non-interacting magnons. After Fourier transformation, $\mathcal{H}_2=\sum_{\bf k} \psi_{\bf k}^{\dagger} \mathcal{H}_{2,{\bf k}} \psi_{\bf k}$, where $\bf k$ is the momentum vector, $\psi_{\bf k}^{\dagger}=(a_{\bf k},b_{\bf k}^{\dagger})$ and
\begin{equation}\label{Eq2}
\mathcal{H}_{2,\bf k}=S\left(
    \begin{array}{cc}
      J_{1}& J_0\\
      J_0 & J_{1}\\
    \end{array}
  \right),
\end{equation}
with $J_{1}=2J\big[\cos(k_x a)+\cos(k_y a)-4\big]+J_0-2K$.

We diagonalize $\mathcal{H}_{2,\bf k}$ via the Bogoliubov transformation \cite{Shen2020}
\begin{equation}\label{Eq3}
\begin{aligned}
a_{\bf k}&=u_{\bf k}\alpha_{\bf k}+v_{\bf k}\beta_{-\bf k}^\dag,\\
b_{-\bf k}&=u_{\bf k}\beta_{-\bf k}+v_{\bf k}\alpha_{\bf k}^\dag,\\
\end{aligned}
\end{equation}
where $u_{\bf k}=\sqrt{(\Delta_{\bf k}+1)/2}$ and $v_{\bf k}=-\sqrt{(\Delta_{\bf k}-1)/2}$, with
\begin{equation}\label{Eq4}
\frac{1}{\Delta_{\bf k}}=\sqrt{1-\Big\{\frac{J_0}{2J[\cos(k_xa)+\cos(k_ya)]+J_0-4J-2K}\Big\}^2}.
\end{equation} This yields the noninteracting magnon energies for branches $\alpha$ and $\beta$: $\varepsilon_{\alpha(\beta)}({\bf k})=S\Big \{2K-2J\big[\cos(k_xa)+\cos(k_ya)-2\big]\Big \}^{1/2}\Big \{2K-2J_0-2J\big[\cos(k_xa)+\cos(k_ya)-2\big]\Big \}^{1/2}$. In this approximation, DMI contributions are absent, rendering the energy dispersions for magnons of opposite chiralities, i.e., right-handed (RH) for $\alpha$ branch and left-handed (LH) for $\beta$ branch, degenerate and concealing any chiral splitting. Capturing the altermagnetic chiral splitting requires incorporating the nonlinear terms $\mathcal{H}_3$ and $\mathcal{H}_4$, which account for magnon-magnon interactions.

\textit{Many-body perturbation theory---}At absolute zero temperature, the quartic term $\mathcal{H}_4$, governing four-magnon interactions, is negligible, as it preserves the symmetries as the harmonic term $\mathcal{H}_2$ and thus does not significantly renormalize the magnon spectrum \cite{Pershoguba2018}. Instead, we focus on the cubic term $\mathcal{H}_3$, mediated by the DMI, which drives three-magnon interactions pivotal to magnon dynamics \cite{SM}. This term reads
\begin{equation}\label{Eq5}
\begin{aligned}
\mathcal{H}_{\rm 3}=\sum_{\bf k, q ,p}^{\bf p=k+q} (V^{11\leftarrow 1}_{\bf k,q\leftarrow p} a_{\bf k}^{\dagger}a_{\bf q}^{\dagger}a_{\bf p}+V^{22\leftarrow 2}_{\bf -k,-q\leftarrow -p}b_{-\bf k}^{\dagger}b_{-\bf q}^{\dagger}b_{-\bf p}+{\rm H.c.}),
\end{aligned}
\end{equation}
where the interaction vertices are $V^{11\leftarrow 1}_{\bf k,q\leftarrow p}=i\sqrt{\frac{2S}{N}}\sum_{\bm\delta}D_{{\rm A},\bm\delta}^{-}(\sin{\bf k}\cdot{\bm\delta}+\sin{\bf q}\cdot{\bm\delta})$ and ${V^{22\leftarrow 2}_{\bf-k,-q\leftarrow -p}}=i\sqrt{\frac{2S}{N}}\sum_{\bm\delta}D_{{\rm B},\bm\delta}^{+}(\sin{\bf k}\cdot{\bm\delta}+\sin{\bf q}\cdot{\bm\delta})$, with $D_{{\rm A}/{\rm B},\bm\delta}^{\pm}={\bf D}_{{\rm A}/{\rm B},\bm\delta}\cdot({\bf e}_y\pm i{\bf e}_x)$. These vertices encapsulate processes unique to bosons, where particle number is not conserved: a single magnon can decay into two, or vice versa, which are prohibited in low-energy fermionic systems.

Rewriting $\mathcal{H}_3$ in the Bogoliubov eigenbasis yields the decay term $\mathcal{H}_{3d}$
\begin{equation}\label{Eq6}
\begin{aligned}
\mathcal{H}_{3d} &=\frac{1}{2}\sum_{\bf k, q ,p}^{\bf p=k+q}\sum_{\lambda \mu \nu}\big[ \mathcal{M}^{\lambda \mu \leftarrow \nu}_{\bf k,q\leftarrow p} c_{\lambda, {\bf k}}^\dagger c_{\mu, {\bf q}}^\dagger c_{\nu, {\bf p}}+ {\rm H.c.}\big]\\
&+\frac{1}{2}\sum_{\bf k, q}\sum_{\lambda \mu \nu}\big[ \mathcal{M}^{\lambda \mu \leftarrow \nu}_{\bf k-q,q\leftarrow k} c_{\lambda, {\bf k-q}}^\dagger c_{\mu, {\bf q}}^\dagger c_{\nu, {\bf k}}+ {\rm H.c.}\big],
\end{aligned}
\end{equation}
and the source term $\mathcal{H}_{3s}$
\begin{equation}\label{Eq7}
\begin{aligned}
\mathcal{H}_{3s} &=\frac{1}{6}\sum_{\bf k, q ,p}^{\bf p=k+q}\sum_{\lambda \mu \nu}\big[ \mathcal{M}^{\lambda \mu \nu}_{{\bf k},{\bf q},{-\bf p}} c_{\lambda,{\bf k}}^\dagger c_{\mu,{\bf q}}^\dagger c_{\nu,{-\bf p}}^\dagger+ {\rm H.c.}\big].
\end{aligned}
\end{equation}
Here, $\mathcal{H}_{3d}$ describes magnon splitting or confluence, while $\mathcal{H}_{3s}$ governs the creation or annihilation of three magnons from vacuum. The operators $c_{\lambda,\mathbf{k}}^{\dagger}$ create magnons in eigenstates ($\lambda = 1$ for $\alpha_{\mathbf{k}}^{\dagger}$, $\lambda = 2$
 for $\beta_{\mathbf{k}}^{\dagger}$), and the vertices $\mathcal{M}^{\lambda \mu \leftarrow \nu}$ are detailed in Sec. III of Supplement Material \cite{SM}. The arrow in $\mathcal{M}^{\lambda \mu \leftarrow \nu}$ indicates the decay of a single magnon into two magnons. To quantify how $\mathcal{H}_3$ renormalizes the magnon spectrum, we apply many-body perturbation theory, computing the single-particle Green's function to order $\frac{1}{S}$ \cite{Mook2021}
\begin{equation}\label{Eq8}
\begin{aligned}
G_{{\bf k},\nu\nu'}(\tau)&=G_{{\bf k},\nu\nu'}^{(0)}(\tau)+\int_0^\beta d{\tau}_1\langle {\mathcal T}{\mathcal{H}_4{\tau_1}c_{\nu,{\bf k}}c_{\nu',{\bf k}}^{\dagger}}\rangle^{\rm con}_{(0)}\\
&-\frac{1}{2}\int_0^\beta d{\tau_1}\int_0^\beta d{\tau_2}\langle {\mathcal T} \mathcal{H}_{3}(\tau_1)\mathcal{H}_{3} (\tau_2) c_{\nu,{\bf k}}(\tau)c_{\nu',{\bf k}}^{\dagger} \rangle^{\rm con}_{(0)},
\end{aligned}
\end{equation}
where ${\mathcal T}$ orders events in imaginary time $\tau$, $\langle \rangle^{\rm con}_{(0)}$ denotes connected averages with respect to $ \mathcal{H}_2$, $\beta=(k_{B}\mathscr{T})^{-1}$ is the inverse temperature ($k_B$: Boltzmann constant, $\mathscr{T}$: temperature), and $G_{{\bf k},\nu\nu'}^{(0)}(\tau)$ is the non-interacting Green's function. 

Applying Wick's theorem leads to three contributing Feynman diagrams, i.e., forward, backward, and circle bubbles \cite{Mook2021}, resulting in the self-energy
\begin{equation}\label{Eq9}
\begin{aligned}
\Sigma_{\mu\mu'}(\varepsilon)&=-\frac{1}{N}\sum_{{\bf q}={\rm BZ}}\sum_{\gamma\gamma'}\bigg\{ \frac{\mathcal{M}_{\bf k,q\leftarrow p}^{\mu \gamma\leftarrow \gamma'*} \mathcal{M}_{\bf k,q\leftarrow p}^{\mu' \gamma \leftarrow \gamma'}}{i\varepsilon-\varepsilon_{\bf p}+\varepsilon_{\bf q}}\big[\rho(\varepsilon_{\bf p})-\rho(\varepsilon_{\bf q})\big]\\
&-\frac{1}{2}\frac{\mathcal{M}_{\bf k-q,q\leftarrow k}^{\gamma \gamma'\leftarrow \nu*} \mathcal{M}_{\bf k-q,q\leftarrow k}^{\gamma \gamma'\leftarrow \nu'}}{\varepsilon-\varepsilon_{\bf q}-\varepsilon_{\bf k-q}}\big[\rho(\varepsilon_{\bf k-q})+\rho(\varepsilon_{\bf q})+1\big]\\
&+\frac{1}{2}\frac{\mathcal{M}_{\bf k,q,-p}^{\nu \gamma \gamma'*} \mathcal{M}_{\bf k,q,-p}^{\nu' \gamma \gamma'}}{\varepsilon+\varepsilon_{\bf q}+\varepsilon_{\bf-p}}\big[\rho(\varepsilon_{\bf q})+\rho(\varepsilon_{\bf-p})+1\big]\bigg\},\\
\end{aligned}
\end{equation}
where $N$ is the number of unit cells, $\rho(\varepsilon)=[\exp(\beta\varepsilon)-1]^{-1}$ is the Bose-Einstein distribution, and the sum runs over the Brillouin zone (BZ) \cite{SM}. The Dyson equation \cite{Mahan2020}
\begin{equation}\label{Eq10}
\begin{aligned}
G_{{\bf k},\nu\nu' }(\varepsilon)=G_{{\bf k},\nu\nu'}^{(0)}(\varepsilon)+\sum_{\mu \mu'}G_{{\bf k},\nu \mu}^{(0)}(\varepsilon)\Sigma_{\mu \mu'}(\varepsilon) G_{{\bf k},\mu' \nu'}(\varepsilon),
\end{aligned}
\end{equation}
then yields the renormalized Hamiltonian matrix
\begin{equation}\label{Eq11}
\mathcal{H}=\left(
    \begin{array}{cc}
      \varepsilon_{\alpha}+\Sigma_{11}(\varepsilon) & \Sigma_{12}(\varepsilon) \\
      \Sigma_{21}(\varepsilon) & \varepsilon_{\beta}+\Sigma_{22}(\varepsilon) \\
    \end{array}
  \right),
\end{equation}
where $\varepsilon_\alpha$ and $\varepsilon_\beta$ are the bare magnon energies. The eigenvalues $\varepsilon'=\varepsilon_{\alpha(\beta)}'$ of Eq. \eqref{Eq11} are given by Eqs. (S33) \cite{SM}. The off-diagonal terms $\Sigma_{12(21)}$ capture spin-flipping scattering, while the diagonal terms $\Sigma_{11(22)}$ shift the energies and can be approximated as
\begin{equation}\label{Eq12}
\begin{aligned}
\Sigma_{11(22)}(\varepsilon)&\approx n_1\Big[\sum_{\bm\delta}{\bf D}_{A(B),\bm\delta}({\bf k}\cdot{\bm \delta})\Big]^2+n_2\Big[\sum_{\bm\delta}{\bf D}_{B(A),\bm\delta}({\bf k}\cdot{\bm \delta})\Big]^2\\
&+n_3\sum_{\bm\delta}\sum_{\bm\delta'}{\bf D}_{A,\bm\delta}\cdot{\bf D}_{B,\bm\delta'}({\bf k}\cdot{\bm \delta})({\bf k}\cdot{\bm \delta'}),\\
\end{aligned}
\end{equation}
with coefficients $n_1$, $n_2$, and $n_3$ depending on magnon energies and Bogoliubov factors $u_{\bf k}$ and $v_{\bf k}$ (see Sec. IV \cite{SM}). Assuming negligible off-diagonal terms and $\varepsilon_{\alpha}=\varepsilon_{\beta}=\varepsilon$, the renormalized energies are reduced to $\varepsilon_{\alpha}'\approx\varepsilon+\Sigma_{11}(\varepsilon),\ \varepsilon_{\beta}'\approx\varepsilon+\Sigma_{22}(\varepsilon)$. If the DMI vectors are identical across sublattices, $\Sigma_{11} = \Sigma_{22}$, yielding uniform band shifts. However, sublattice-dependent DMIs lead to $\Sigma_{11} \neq \Sigma_{22}$, inducing magnon chiral splitting.

\begin{figure}[htbp]
  \centering
  \includegraphics[width=0.48\textwidth]{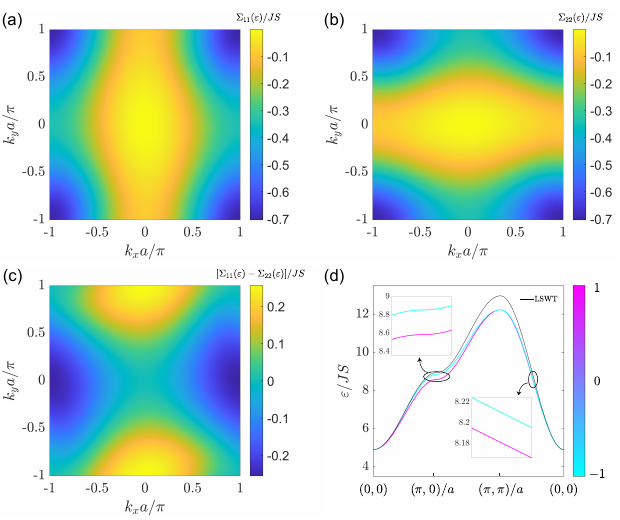}\\
  \caption{Momentum-space distributions of self-energy components $\Sigma_{11}(\varepsilon)$ (a) and $\Sigma_{22}(\varepsilon)$ (b) for the $H^{\rm C}$ model. (c) Quadrupole-like chiral splitting $\Sigma_{11}(\varepsilon)$-$\Sigma_{22}(\varepsilon)$ manifesting $C_4 T$ symmetry. (d) Renormalized magnon spectrum from Eq. \eqref{Eq11} (gradient-colored curves) compared to LSWT predictions (black lines). The inset offers a zoomed-in view. The color bar denotes the average magnon chirality $P$. Parameters: $\mathscr{T}=0$ K, ${\bf D}_{A,x}=0.3J(\frac{\sqrt{2}}{2},\frac{\sqrt{2}}{2})$, ${\bf D}_{A,y}=0.15J(0,1)$, ${\bf D}_{B,x}=0.15J(1,0)$, and ${\bf D}_{B,y}=0.3J(-\frac{\sqrt{2}}{2},\frac{\sqrt{2}}{2})$.}\label{fig2}
\end{figure}

\textit{Altermagnetic chiral splitting effect---}We adopt the $C$-type DMI as a representative case. This interaction is characterized by specific relations between components across sublattices: $D_{A,xx}=D_{B,yy}$, $D_{A,xy}=-D_{B,yx}$, $D_{A,yx}=-D_{B,xy}$, and $D_{A,yy}=D_{B,xx}$, where three subscripts respectively denote the sublattice (A or B), the direction of the NN bond (e.g., $x$ or $y$) and the component of the DMI vector (also $x$ or $y$). In conventional magnetic systems, DMI vectors are typically perpendicular to the bonds, preserving mirror symmetries. However, in this $C$-type configuration, they deviate from such alignment, violating crystallographic mirror symmetry and yielding asymmetric self-energy contributions $\Sigma_{11(22)}(\varepsilon)$, as evident in Figs. \ref{fig2}(a) and (b). The resultant chiral splitting differentiates the energies of magnons with opposite chiralities, quantified by
\begin{equation}\label{Eq14}
\begin{aligned}
\Sigma_{11}(\varepsilon)-\Sigma_{22}(\varepsilon)&\approx \\
&w_{1}({\bf D}_{A,x}^2-{\bf D}_{A,y}^2)(k_x^2-k_y^2)+4w_{1}{\bf D}_{A,x}\cdot{\bf D}_{A,y}k_xk_y,
\end{aligned}
\end{equation}
where $w_1 = n_1 - n_2$ \cite{SM}. This expression reveals a quadrupole-like dependence on wavevector $\mathbf{k}$, with terms proportional to $k_x^2 - k_y^2$ and $k_x k_y$, as corroborated by our anharmonic theory in Fig. \ref{fig2}(c). Figure \ref{fig2}(d) illustrates the renormalized magnon spectrum from Eq. \eqref{Eq11}; along paths from $(0,0)$ to $(\pi/a,0)$ and $(0,0)$ to $(\pi/a,\pi/a)$, the splitting endures despite effective magnon SOC originating from terms $\Sigma_{12(21)}(\varepsilon)$. To quantify the average magnon chirality in the two split bands, we define the polarization $P=(|c_{\alpha}|^2-|c_{\beta}|^2)/(|c_{\alpha}|^2+|c_{\beta}|^2)$ for each renormalized magnon eigenstate $c_{\alpha}\alpha_{\bf k}+c_{\beta}\beta_{\bf k}$ obtained by diagonalizing the Hamiltonian \eqref{Eq11} \cite{Li2024}. Here, the coefficients $c_{\alpha}=\big[\varepsilon'-\varepsilon-\Sigma_{22}(\varepsilon)\big]/\Sigma_{12}(\varepsilon)$ and $c_{\beta}=1$ represent the amplitudes in the bare magnon basis $\alpha_{\bf k}$ and $\beta_{\bf k}$, respectively. Numerical results for $P$ are visualized via color gradients along the dispersion curves in Fig. \ref{fig2}(d). This framework can extend beyond $C$-type DMI. Applying similar procedure, we find $\Sigma_{11}(\varepsilon)-\Sigma_{22}(\varepsilon)\propto{\bf D}_{A,x}\cdot{\bf D}_{A,y}k_xk_y$ for $M$-type DMI, and $\Sigma_{11}(\varepsilon)-\Sigma_{22}(\varepsilon)\propto{(\bf D}_{A,x}^2-{\bf D}_{A,y}^2)(k_x^2-k_y^2)$ for $CM$-type DMI, with comprehensive details in Sec. V \cite{SM}. Such adaptability underscores our method's efficacy in addressing interaction-driven altermagnetic chiral splitting under diverse symmetries.

\textit{Discussion---}The DMI can hybridize magnons with opposite chiralities via the off-diagonal self-energy terms $\Sigma_{12(21)}(\varepsilon)$. In our bilayer model, the dominant interaction terms $\mathcal{M}^{\lambda \mu \leftarrow v}_{\mathbf{k},\mathbf{q} \leftarrow \mathbf{p}}$ primarily affect the diagonal self-energy components $\Sigma_{11(22)}$, which renormalize the magnon energies, while exerting only a negligible influence on the off-diagonal terms \cite{SM}. Consequently, the off-diagonal contributions remain substantially weaker than the altermagnetic chiral splitting (two orders of magnitude smaller), thereby preserving the hallmark altermagnetic features. At elevated temperatures, thermal fluctuations increase the magnon population, thereby enhancing the self-energy corrections and amplifying the chiral splitting according to the power law $(\Sigma_{11}-\Sigma_{22})/JS\propto(k_{B}\mathscr{T}/JS)^{8}$ superimposed on the zero-temperature contribution (see Fig. S4 in Sec. VI \cite{SM}). This unusually high exponent stems from dominant high-momentum contributions in the DMI-dependent three-magnon vertices under thermal activation, potentially constituting a distinctive hallmark of interaction-driven altermagnetism. Furthermore, the strength of three-magnon interactions is scaled as $S^{-3/2}$, meaning that they are particularly influential for small spin lengths, whereas for larger spins ($S \geq 5$), the magnon spectrum converges to that predicted by LSWT (see Sec. VII \cite{SM}).

In this study, the DMI manifests in three distinct forms, each linked to specific symmetries, which can be tailored by modulating the SOC \cite{Gungordu2016}. For systems preserving both $C_4 T$ and $\sigma_v T$ symmetries, combining Rashba and Dresselhaus SOC yields $\mathbf{D}_{A,x} = \mathbf{D}_{B,y} = (0, D_R + D_D)$ and $\mathbf{D}_{A,y} = \mathbf{D}_{B,x} = (-D_R + D_D, 0)$; In the case of $\sigma_v T$ symmetry alone, a nonorthogonal DMI emerges as $\mathbf{D}_{A,x} = (-D_D, D_R)$, $\mathbf{D}_{A,y} = (-D_R, D_D)$, $\mathbf{D}_{B,x} = (D_D, D_R)$, and $\mathbf{D}_{B,y} = (-D_R, -D_D)$; For $C_4 T$ symmetry only, anisotropic DMI driven by Dresselhaus SOC produces $\mathbf{D}_{A,x} = (-D_D, D_1)$, $\mathbf{D}_{A,y} = (-D_2, D_D)$, $\mathbf{D}_{B,x} = (D_D, D_2)$, and $\mathbf{D}_{B,y} = (-D_1, -D_D)$ \cite{Gungordu2016}. Here, $D_R$ and $D_D$ denote the DMI strength from Rashba and Dresselhaus SOC, respectively, while $D_1$ and $D_2$ account for unequal strengths in anisotropic DMI. These configurations are realized in materials such as noncentrosymmetric Heusler compounds \cite{Peng2020} and epitaxial Au/Co/W(110) films \cite{Camosi2017}. The predicted chiral splitting, up to $0.02\varepsilon'$, is detectable via inelastic neutron scattering \cite{Lovesey1977}, as recently demonstrated in $\alpha$-MnTe \cite{Liu2024}. 

To conclude, our study unveils a novel mechanism in compensated magnets: in the relativistic regime, nonlinear magnon-magnon interactions can induce significant altermagnetic chiral splitting. Through detailed analysis of the cubic (three-magnon) interaction terms arising from the DMI, we identify three distinct classes of chiral splitting. Each is governed by specific symmetries, i.e., combinations of rotation, reflection, and time reversal, yielding unique splitting patterns overlooked by linear harmonic approximations. Central to our discovery is a bosonic mechanism that drives this symmetry-protected chiral splitting, exploiting magnons' inherent violation of particle-number conservation, a feature absent in low-energy electron systems. Furthermore, the unusually high temperature scaling of the splitting $(\propto \mathscr{T}^8)$ implies exceptional robustness against thermal fluctuations at low temperatures, potentially serving as a distinctive experimental signature of interaction-driven altermagnetism. We anticipate that this predicted interaction-induced magnon chiral splitting may emerge in a broader range of anisotropic antiferromagnets. Our findings illuminate altermagnetic order in bosonic contexts, deepen our understanding of quantum materials, and pave the way for innovative advancements in magnonics and allied fields.
\begin{acknowledgments}
This work was funded by the National Key R$\&$D Program under Contract No. 2022YFA1402802, the National Natural Science Foundation of China (NSFC) (Grants No. 12374103 and No. 12434003), and Sichuan Science and Technology Program (No. 2025NSFJQ0045). Z. J. acknowledges the financial support from NSFC (Grant No. 12404125) and the China Postdoctoral Science Foundation (Grant No. 2024M750337). Y. S. was supported by Sichuan Science and Technology Program (Grant No. 2025ZNSFSC0866).
\end{acknowledgments}

\end{document}